\documentclass[12pt]{article}
\usepackage{amsfonts,epsfig,psfig,cite}

\begin{document}

\date{February 26, 2002}

\def\reff#1{(\ref{#1})}
\newcommand{\be}{\begin{equation}}
\newcommand{\ee}{\end{equation}}
\newcommand{\<}{\langle}
\renewcommand{\>}{\rangle}

\def\spose#1{\hbox to 0pt{#1\hss}}
\def\ltapprox{\mathrel{\spose{\lower 3pt\hbox{$\mathchar"218$}}
 \raise 2.0pt\hbox{$\mathchar"13C$}}}
\def\gtapprox{\mathrel{\spose{\lower 3pt\hbox{$\mathchar"218$}}
 \raise 2.0pt\hbox{$\mathchar"13E$}}}

\def\bsigma{\mbox{\protect\boldmath $\sigma$}}
\def\bpi{\mbox{\protect\boldmath $\pi$}}
\def\smfrac#1#2{{\textstyle\frac{#1}{#2}}}
\def\smhalf{ {\smfrac{1}{2}} }

\newcommand{\re}{\mathop{\rm Re}\nolimits}
\newcommand{\im}{\mathop{\rm Im}\nolimits}
\newcommand{\tr}{\mathop{\rm tr}\nolimits}

\def\Z{{\mathbb Z}}
\def\R{{\mathbb R}}
\def\C{{\mathbb C}}

\title{Two-Dimensional Heisenberg Model with Nonlinear Interactions}
\author{
  \\[-0.5cm]
  {\small Sergio Caracciolo}              \\[-0.2cm]
  {\small\it Dipartimento di Fisica dell'Universit\`a di Milano, I-20100 Milano}
                     \\[-0.2cm]
  {\small\it INFN, Sez. di Pisa, and INFM-NEST, Pisa, ITALIA}    \\[-0.2cm]
  {\small {\tt Sergio.Caracciolo@sns.it}}     \\[-0.2cm]
  \\[-0.35cm] \and
  {\small Andrea Pelissetto }        \\[-0.2cm]
  {\small\it Dipartimento di Fisica and INFN -- Sezione di Roma I}  \\[-0.2cm]
  {\small\it Universit\`a degli Studi di Roma ``La Sapienza''}      \\[-0.2cm]
  {\small\it I-00185 Roma, ITALIA}        \\[-0.2cm]
  {\small {\tt Andrea.Pelissetto@roma1.infn.it}}  \\[-0.2cm]
  {\protect\makebox[5in]{\quad}}  
   \\
}

\vspace{0.2cm}
\maketitle
\thispagestyle{empty}   

\vspace{-0.4cm}

\begin{abstract}
We investigate a two-dimensional classical $N$-vector model 
with a nonlinear interaction $(1 + \bsigma_i\cdot \bsigma_j)^p$
in the large-$N$ limit. As observed for $N=3$ by 
Bl\"ote {\em et al.} [Phys. Rev. Lett. {\bf 88}, 047203 (2002)],
we find a first-order transition for $p>p_c$ and no finite-temperature
phase transitions for $p < p_c$.
For $p>p_c$, both phases have short-range order, the correlation length
showing a finite discontinuity at the transition.
For $p=p_c$, there is a peculiar transition, where the spin-spin
correlation length is finite while the energy-energy correlation length
diverges.

\bigskip 

PACS: 75.10.Hk, 05.50.+q, 64.60.Cn, 64.60.Fr

\end{abstract}

\clearpage


The two-dimensional Heisenberg model has been the object of 
extensive studies which mainly focused on the $O(N)$-symmetric
Hamiltonian
\be
H = - N \beta \sum_{\<ij\>} \bsigma_i\cdot \bsigma_j,
\ee
where $\bsigma_i$ is an $N$-dimensional unit spin and 
the sum is extended over all lattice nearest neighbors.
The behavior of this system in two dimensions is well
understood. It is disordered for all finite $\beta$
\cite{MW-66} and it is described for $\beta\to\infty$ 
by the perturbative renormalization group 
\cite{Polyakov-75,BZ-76,BLS-76}. The square-lattice
model has been extensively studied numerically
\cite{Wolff_89_90,EFGS-92,Kim,CEPS-95,CEMPS-96,MPS-96},
checking the perturbative predictions 
\cite{FT-86,CP-94,CP-95,ACPP-99,ABC-97}
and the nonperturbative constants 
\cite{HMN-90,HN-90,CPRV-97}.

In this paper we study a more general Hamiltonian 
on the square lattice; more precisely, we consider
\be
H = - N\beta \sum_{x\mu} W(1 + \bsigma_x\cdot \bsigma_{x+\mu}),
\label{Hgenerica}
\ee
where $W(x)$ is a generic function such that 
$W(2) > W(x)$ for all $0\le x < 2$, in order to
guarantee that the system orders ferromagnetically 
for $\beta\to\infty$. A particular case of 
the Hamiltonian \reff{Hgenerica} has been extensively
studied in the years, the case in which $W(x)$ is 
a second-order polynomial. Such a choice of $W(x)$
gives rise to the so-called mixed
$O(N)$-$\R P^{N-1}$ model 
\cite{HM-82,MR-87,Ohno_90,KZ,%
Butera_92,CEPS_RP,Hasenbusch_96,%
Niedermayer_96,Catterall_98,SS-01}, which is relevant 
for liquid crystals \cite{Lasher-72,LL-72,KS-87,Lin-87,LL-87,BZCP-91}
and for some orientational transitions \cite{KJ-54}.

In a recent Letter \cite{BGH-02}, the authors analyzed 
a model with $W(x)= a x^p + b$ and found an additional
first-order transition for $p$ large enough. 
Here, we will study the same model, finding an 
analogous result: for $p>p_c\approx 4.537857$ a first-order
transition appears, the correlation length---and 
in general, all thermodynamic quantities---showing 
a finite discontinuity. Note that the appearance of a 
first-order transition in nonlinear models is not a new 
phenomenon. Indeed, for $N=\infty$ it was already
shown in Ref. \cite{MR-87} that a first-order transition
appears in mixed $O(N)$-$\R P^{N-1}$ models for some 
values of the couplings. It is of interest to understand the 
behavior for $p=p_c$. For such value of $p$,
Ref. \cite{BGH-02} found a peculiar phase transition:
while the spin-spin correlation length remains finite, the 
energy-energy correlation length diverges. 
Here, we will show that the same phenomenon occurs for $N=\infty$.
However, at variance with what observed in Ref. \cite{BGH-02},
the critical theory shows mean-field---not Ising---behavior.

Let us consider the Hamiltonian \reff{Hgenerica}
on a hypercubic $d$-dimensional lattice. We normalize 
$W(x)$ by requiring $W'(2)=1$ so that in the spin-wave limit 
\be
H = {N\beta\over 2} \int dx\, \partial_\mu\bsigma\cdot \partial_\mu\bsigma.
\ee 
We also fix 
$W(1) = 0$ so that $H=0$ for a random configuration.
Then, we introduce two new fields 
$\lambda_{x\mu}$ and $\rho_{x\mu}$ in order to linearize the 
dependence of the Hamiltonian on the spin coupling. We write
\be
\exp\left[ N\beta\, W(1 + \bsigma_x\cdot\bsigma_{x+\mu})\right] \sim
\int d\rho_{x\mu} d \lambda_{x\mu} \exp\left[
  {N\beta\over2} \lambda_{x\mu} 
      \left(1 + \bsigma_x\cdot\bsigma_{x+\mu} - \rho_{x\mu}\right)
  + N \beta W(\rho_{x\mu})\right].
\ee
As usual in the large-$N$ expansion, we also introduce a field $\mu_x$
in order to eliminate the constraint $\bsigma_x^2 = 1$. Thus, we write
\be
\delta\left(\bsigma_x^2 - 1\right) \sim 
\int d\mu_x \exp\left[
  - {N\beta\over2} \mu_x\left(\bsigma_x^2 - 1\right)\right].
\ee
With these transformations we can rewrite the partition function as 
\be
Z = \int \prod_{x\mu} [d\rho_{x\mu} d \lambda_{x\mu}]
         \prod_x [d\mu_x d\bsigma_x]\,  e^{NA}
\ee
where 
\be
A = {\beta\over2} \sum_{x\mu} 
   \left[\lambda_{x\mu} + \lambda_{x\mu}\bsigma_x\cdot\bsigma_{x+\mu} - 
         \lambda_{x\mu} \rho_{x\mu} + 2 W(\rho_{x\mu})\right] - 
   {\beta\over2} \sum_{x} 
     \left(\mu_x\bsigma_x^2 - \mu_x\right).
\ee
We perform a saddle-point integration by writing
\begin{eqnarray}
\lambda_{x\mu} &=& \alpha + \widehat{\lambda}_{x\mu}, \nonumber \\
\rho_{x\mu} &=& \tau + \widehat{\rho}_{x\mu}, \nonumber \\
\mu_{x} &=& \gamma + \widehat{\mu}_{x}. 
\end{eqnarray}
A standard calculation gives the following saddle-point equations
\cite{foot1}:
\begin{eqnarray}
&& d\beta(1-\tau) + {1\over \alpha} \left[(2 d + m_0^2) I(m_0^2) - 1\right] = 0,
\nonumber \\
&& \alpha - 2 W'(\tau) = 0, 
\nonumber \\
&& {\beta\over2} - {1\over \alpha} I(m_0^2) = 0,
\label{puntisella}
\end{eqnarray}
where we set $\gamma = \alpha(2d + m_0^2)/2$,
\be
I(m_0^2) = \int{d^dp\over (2\pi)^d}\, {1\over \hat{p}^2 + m_0^2},
\ee
and $\hat{p}^2 = 4 \sum_\mu \sin^2 p_\mu/2$. The variable $m_0$
has a simple interpretation: it is related to the 
spin-spin correlation length by $\xi_\sigma = 1/m_0$. 
From Eq. \reff{puntisella} we obtain finally
\be
\beta = {I(m_0^2)\over W'(\tau)},
\label{puntosella2}
\ee
where 
\be
\tau = \tau(m_0) \equiv 2 + {m_0^2\over 2d} - {1\over 2 d I(m_0^2)}.
\ee
The corresponding free energy can be written as 
\be
F = - \beta d W(\tau) + {1\over 2} \log I(m_0^2) + {1\over2} L(m_0^2),
\ee
where 
\be
L(m_0^2) = \int{d^dp\over (2\pi)^d}\,\log (\hat{p}^2 + m_0^2).
\ee
Focusing now on the two-dimensional case, let us show that, for any
$W(x)$, the spin-spin correlation length is always finite, i.e.  
$\xi_\sigma=\infty$, i.e. $m_0=0$, only for $\beta=\infty$.
Note first that $\tau = 2$ (resp. $\tau = 1$) for 
$m_0 = 0$ (resp. $m_0 = \infty$) and that $\tau(m_0)$ is 
a strictly decreasing function of $m_0$. Thus, 
$W'(\tau)$ is finite for all $m_0$. 
Then, since $I(0) = +\infty$, we find that $\xi_\sigma=\infty$  
only if $\beta=\infty$,
i.e. $\xi_\sigma$ is finite for all
finite $\beta$. 

\begin{figure}[tb]
\hspace{-1cm}
\vspace{0cm}
\centerline{\psfig{width=12truecm,angle=-90,file=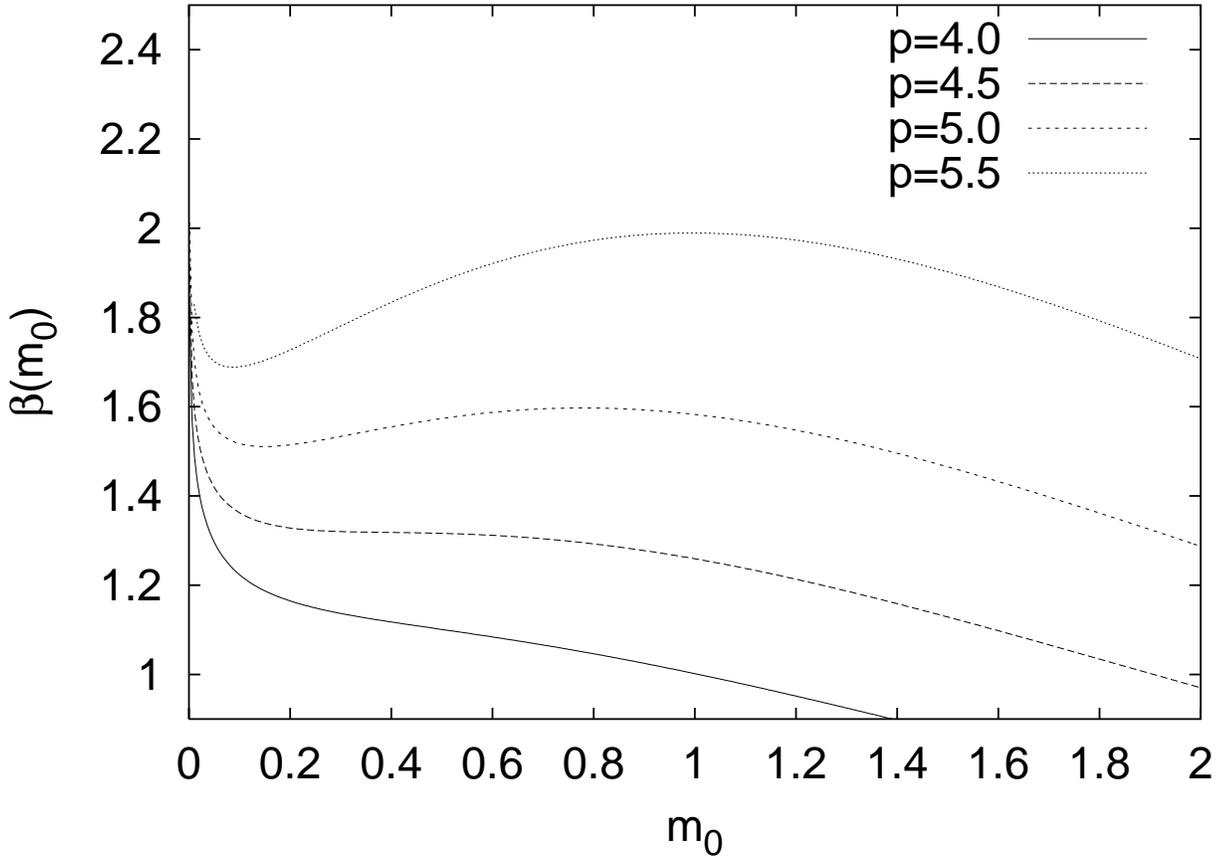}}
\vspace{0cm}
\caption{Function $\beta(m_0) \equiv I(m_0)/W'(\tau)$ vs $m_0$, for 
$p=4$, 4.5, 5, and 5.5. 
For any $p$, $\beta(m_0)\to\infty$ for $m_0\to 0$.}
\label{fig-beta}
\end{figure}

We want now to discuss the behavior for $\beta\to\infty$. 
From Eq. \reff{puntosella2}, we see that $\beta\to\infty$ for 
$m_0\to 0$ and possibly for $m_0\to \bar{m}_i$, where 
$W'(\tau(\bar{m}_i))=0$. If there is more than one solution, 
the relevant one corresponds to the lowest free energy. 
Now, for $\beta\to\infty$, we can simply write \cite{foot2}
$F \approx - 2 \beta W(\tau)$. Since $\tau(0)=2$ and 
$W(2) > W(\tau)$ for all $0\le \tau < 2$ because of the 
ferromagnetic condition, the relevant solution is the 
one with $m_0\to 0$. Then, using 
\be
I(m_0) = -{1\over2\pi} \log{m_0^2\over32} + O(m_0^2 \log m_0^2)
\ee
for $m_0\to 0$, we obtain 
\be
m_0^2 = 32 e^{-2\pi\beta + \pi W''(2)/2}[1 + O(\beta^{-1})],
\ee
in agreement with the standard perturbative renormalization-group 
predictions \cite{foot3}.

\begin{figure}[tb]
\hspace{-1cm}
\vspace{0cm}
\centerline{\psfig{width=12truecm,angle=-90,file=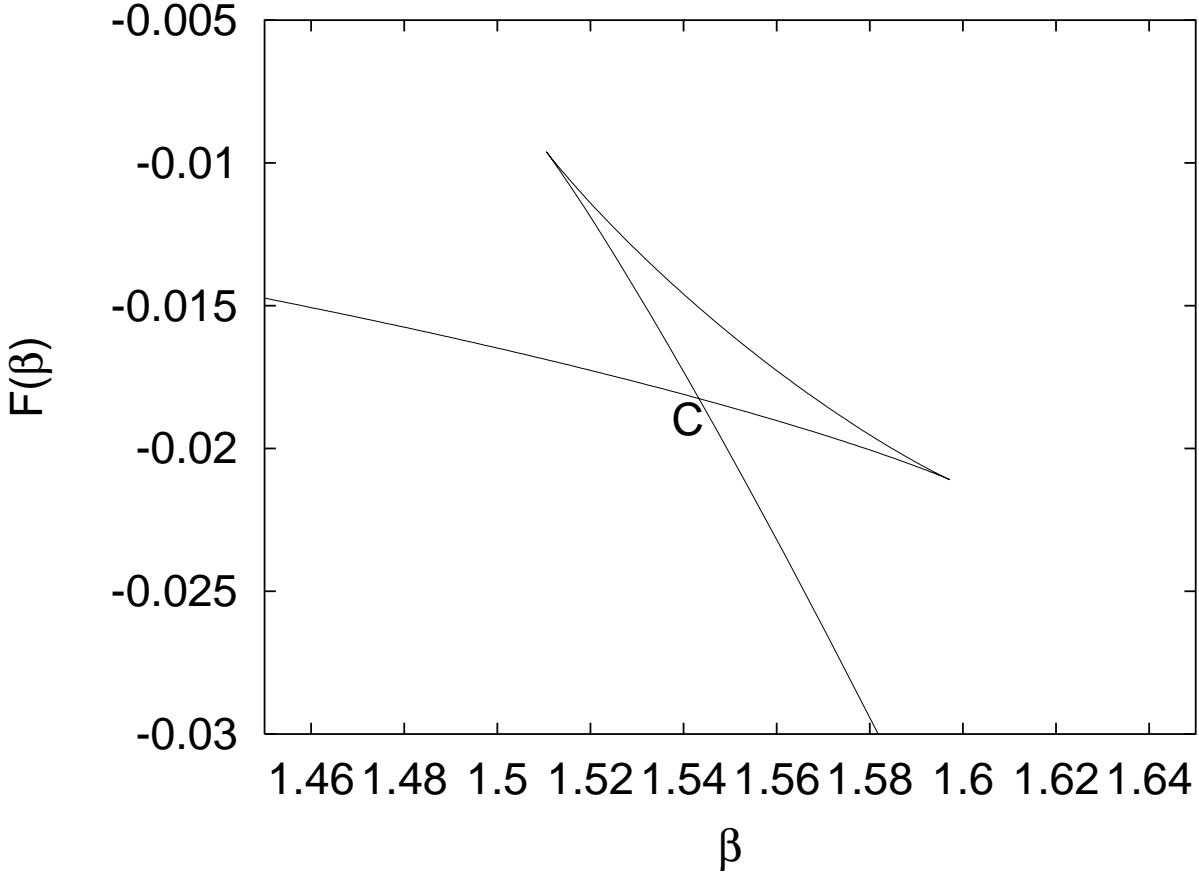}}
\vspace{0cm}
\caption{The free energy $F(\beta)$ for $p=5$.
There is a critical point $C$ for $\beta_c \approx 1.543$.}
\label{fig-F}
\end{figure}

Let us now discuss the possibility of first-order phase transitions, which 
may arise from the presence of multiple solutions to 
Eq. \reff{puntosella2}. As in Ref. \cite{BGH-02}, we consider 
\be
W(x) = {2\over p}\left( {x\over2}\right)^p - {2^{1-p}\over p}.
\ee
In Fig. \ref{fig-beta} we report the function 
$\beta(m_0) \equiv I(m_0^2)/W'(\tau)$, for $p = 4,4.5,5,5.5$.
For $p=4,4.5$, for each $\beta$ there is a unique solution
$m_0$ and thus there are no phase transitions. On the other hand,
for $p=5,5.5$ there is the possibility of multiple solutions, in which case
the most relevant is the one that gives the lowest free energy. 
For $p=5$, we report the free energy in Fig. \ref{fig-F}.
We observe a first-order transition for 
$\beta \approx 1.543$ with a finite discontinuity of the correlation length,
$\Delta \xi_\sigma \approx 16.2$, and of all thermodynamic quantities. 
A numerical analysis of the gap equation
\reff{puntosella2} shows that a first-order transition exists 
for all $p > p_c \approx 4.537857$. For $p=p_c$, the thermodynamic functions
are nonanalytic for $\beta = \beta_c\approx 1.33472$.
In this case, 
\be
\beta - \beta_c \approx -0.035726 (m_0 - m_{0c})^3 + O[(m_0 - m_{0c})^4],
\ee
where $m_{0c} \approx 0.387537$. Consequently, repeating the discussion of 
Ref. \cite{MR-87},
\begin{eqnarray}
\xi_\sigma(\beta) &\approx& 2.5804 + 7.8682 (\beta - \beta_c)^{1/3} + 
      \cdots , \\
E (\beta) &\approx& 0.162274  + 0.314385 (\beta - \beta_c)^{1/3} + \cdots , \\
C (\beta) &\approx& 0.104795 (\beta - \beta_c)^{-2/3} + \cdots,
\end{eqnarray}
where $E$ and $C$ are respectively the energy and the specific heat per site.
Note that $C (\beta)$ diverges at the critical point, indicating that,
although spin-spin correlations are not critical, criticality is 
observed for energy-energy correlations. Indeed, consider
\be
D_Q(k) = \sum_{x\mu\nu} e^{ik\cdot(x-y)} 
   \langle Q(1 + \bsigma_x\cdot \bsigma_{x+\mu});
           Q(1 + \bsigma_y\cdot \bsigma_{y+\nu})\rangle,
\ee
where $Q(x)$ is an arbitrary regular function. For $N\to \infty$, 
\be
D_Q(k) = [Q'(\tau)]^2 \sum_{\mu\nu} 
     \langle \widehat{\rho}_\mu(-k); \widehat{\rho}_\nu(k)\rangle,
\ee 
so that
\be
N D_Q(0) = \left({Q'(\tau)\over W'(\tau)}\right)^2 C(\beta).
\ee
It follows $D_Q(0) \sim (\beta-\beta_c)^{-2/3}$ for any function $Q(x)$.
Thus, all correlation functions of the energy show a critical behavior.
In order to compute the associated correlation length, we determine 
$D_Q(k)$ for arbitrary $k$. We obtain
\be
N D_Q(k) = {2 [Q'(\tau)]^2[ A_2(k) A_0(k) - A_1(k)^2] \over 
   \beta^2 [W'(\tau)]^2 A_0(k) - \beta W''(\tau)[ A_2(k) A_0(k) - A_1(k)^2]},
\ee
where
\be
A_n(k) = \left.\int {d^2q\over (2\pi)^2} \, 
    { \left(\sum_\mu \cos q_\mu\right)^n \over 
    \left( \widehat{(q + k/2)}^2 + m_0^2\right)
    \left(\widehat{(q - k/2)}^2 + m_0^2\right) }\right. \; .
\ee
For $\beta\to \beta_c$ and $k\to 0$, we have 
\be
D_Q(k)^{-1} = a (\beta - \beta_c)^{2/3} + b k^2 + O(k^4),
\ee
with $a,b\not=0$. Thus, the energy-energy correlation length $\xi_E(\beta)$
behaves as 
\be
\xi_E(\beta) \sim (\beta - \beta_c)^{-1/3},
\ee
i.e. $\nu_E = 1/3$. We thus confirm the results of Ref. \cite{BGH-02}
on the existence of the critical theory for $p=p_c$, although we 
disagree on the nature of the critical behavior. Indeed,
Ref. \cite{BGH-02} suggested $\alpha = 1 - 1/\delta$, with $\delta$
assuming the Ising value $\delta=15$. 
Instead, we find the mean-field value $\delta = 3$.
It is unclear how our large-$N$ result is compatible with what observed 
for $N=3$. Indeed, the universality argument of Ref. \cite{BGH-02} 
would predict Ising behavior for any value of $N$. This issue deserves 
further investigations.

\bigskip

We thank Henk Bl\"ote and Henk Hilhorst for many useful comments.


\begin{thebibliography}{199}
\bibitem{foot1}
If $W'(1) = 0$, there is an additional solution with 
$\tau = 1$, $\alpha = 0$, $\gamma = \beta^{-1}$.
\bibitem{foot2}
This is obvious for $m_0\to\bar{m}_i$, since $I(\bar{m}_i)$
and $L(\bar{m}_i)$ are finite. For $m_0\to 0$, it is enough to 
observe that $L(0)$ is finite, while $\log I(m_0^2)\approx \log \beta$
because of the gap equations.
\bibitem{foot3}
The large-$\beta$ behavior of $\xi_\sigma$ for any $N\ge 3$ and any 
potential $W(x)$ is given in Ref.~\cite{CP-94}.
\end{thebibliography}
\end{document}